\documentclass[floatfix,aps,amsmath,nofootinbib,twocolumn,10pt]{revtex4}

\usepackage{listings}
\usepackage{graphicx}
\usepackage{bm}
\usepackage{rotating}
\usepackage{array}
\usepackage{amsmath}
\usepackage{amssymb} 
\usepackage{mathrsfs} 
\usepackage{cancel}
\usepackage{subfig}
\usepackage{float}
\usepackage{caption}
\usepackage{xcolor}
\usepackage{ulem}
\newcommand{\MovB}[1]{{\color{black}{#1}}}
\newcommand{\Mov}[1]{{\color{black}{#1}}}
\newcommand{\MovE}[1]{{\color{black}{#1}}}
\usepackage{multirow}

\def\({\left(}
\def\){\right)}
\def\[{\left[}
\def\]{\right]}

\def\e{\begin{equation}}
\def\q{\end{equation}}
\def\m{\begin{eqnarray}}
\def\n{\end{eqnarray}}

\begin{document}
\title{Constraints on the Local Cosmic Void from the Pantheon Supernovae Data}
\author{Ke Wang$^{1,2,3,4}$}
\thanks{Corresponding author: {wangkey@lzu.edu.cn}}
\author{Kun-Peng Chen$^{4}$}
\thanks{chenkp19@lzu.edu.cn}
\author{Morgan Le Delliou$^{1,2,3,5,6}$}
\thanks{(delliou@lzu.edu.cn,)morgan.ledelliou.ift@gmail.com}
\affiliation{$^1$Institute of Theoretical Physics $\&$ Research Center of Gravitation, Lanzhou University, Lanzhou 730000, China}
\affiliation{$^2$Key Laboratory of Quantum Theory and Applications of MoE, Lanzhou University, Lanzhou 730000, China}
\affiliation{$^3$Lanzhou Center for Theoretical Physics $\&$ Key Laboratory of Theoretical Physics of Gansu Province, Lanzhou University, Lanzhou 730000, China}
\affiliation{$^4$School of Physical Science and Technology, Lanzhou University, Lanzhou 730000, China}
\affiliation{$^5$Instituto de Astrof\'isica e Ci\^encias do Espa\c co, Universidade de Lisboa,
1769-016 Lisboa, Portugal}
\affiliation{$6$Universit\'e de Paris Cit\'e, APC-Astroparticule et Cosmologie (UMR-CNRS 7164), 
 F-75006 Paris, France.}

\date{\today}

\begin{abstract}
In principle, the local cosmic void can be simply modeled by the spherically symmetric Lemaitre-Tolman-Bondi (LTB) metric. In practice, the real local cosmic void is probably not spherically symmetric. In this paper, to reconstruct \Mov{a more }realistic profile of the local cosmic void, we divide it into several segments. Each segment with certain solid angle is modeled by its own LTB metric. Meanwhile, we divide the 1048 type Ia supernovae (SNIa) of \Mov{the Pantheon Survey }into corresponding subsets according to their distribution in the galactic coordinate system. Obviously, each SNIa subset can only be used to reconstruct the profile of one segment. Finally, we can patch together an irregular profile for the local cosmic void with the whole Pantheon sample. \MovB{Note that, the paucity of each data subset lead us to focus on the inner part of each void segment and assume that the half radii of the void segments are sufficient to constrain the whole segment.
We find that, despite $2\sigma$ signals of anisotropy limited to the depth of the void segments, the constraints on every void segment are consistent with $\Lambda$CDM model at $95\%$ CL. 
Moreover, our constraints are too weak to challenge the cosmic homogeneity and isotropy.} 
\end{abstract} 

\maketitle


\section{Introduction}
\label{sec:intro}
The cosmological principle assumes that the universe is homogeneous and isotropic on cosmic scales. Based on this assumption as well as the standard model of particle physics and Einstein's general relativity (GR), the Lambda cold dark matter ($\Lambda$CDM) model is proposed. This standard model of cosmology has proved successful on large cosmic scales according to the latest cosmic microwave background (CMB) observations, namely Planck 2018 data~\cite{Planck:2018vyg}\MovE{, DES \cite{DES:2017myr} or eBOSS \cite{SDSS-IV:2019txh}}. However, it faces major challenges on small scales, such as the Hubble tension~\cite{Riess:2019cxk}\MovE{, the $\sigma_8$ tension \cite{Planck:2015lwi,KiDS:2020ghu,Sakr:2021jya}} and the dipolar tension~\cite{Secrest:2020has}. Besides GR, dark energy model or treatments of systematic uncertainty, the former tension also challenges the local cosmic homogeneity and the latter one challenges \Mov{as well }the cosmic isotropy on small scales. Therefore, further test\Mov{ing} of the local cosmic inhomogeneity and anisotropy on small scales is necessary. 

The local cosmic inhomogeneity can be modeled by the spherically symmetric Lemaitre-Tolman-Bondi (LTB) metric~\cite{Lemaitre:1933gd,Tolman:1934za,Bondi:1947fta}. \Mov{Considering} the late-time matter and dark energy, \Mov{we can use it as an inhomogeneous generalisation of a }$\Lambda$CDM model, namely \Mov{as a }$\Lambda$LTB model. Using the combination of the latest available cosmological observations, the local radial inhomogeneity in the $\Lambda$LTB model has been probed~\cite{Camarena:2021mjr,Camarena:2022iae}. Even though a deeper local void can reconcile the Hubble tension~\cite{Marra:2013rba} and a larger local void can reconcile the dipolar tension~\cite{Cai:2022dov}, a shallower and smaller local void is favored by the combination of the latest available cosmological observations~\cite{Camarena:2021mjr,Camarena:2022iae}. \Mov{However}, a spherically symmetric local cosmic inhomogeneity may not meet reality. Therefore, in this paper, we will probe the true profile of the local cosmic inhomogeneity as realistic\Mov{ally} as possible.

The cosmic anisotropy on small scales can be tested by type Ia supernovae (SNIa) data, such as the combined Pantheon sample~\cite{Pan-STARRS1:2017jku}. One straightforward method is to divide the whole samples into several subsets according to the distribution of individual SNIa in the galactic coordinate system. In particular\Mov{, the} hemisphere comparison method~\cite{Schwarz:2007wf,Antoniou:2010gw} divides the whole sample into two data subsets which are \Mov{designated as the }``up'' and \Mov{the }``down'' hemisphere\Mov{s,} respectively~\cite{Deng:2018yhb,Sun:2018cha,Deng:2018jrp}. \Mov{Furthermore}, $\mathsf{HEALPix}$~\cite{Gorski:2004by} can be used to divide the whole sample into more data subsets~\cite{Deng:2018jrp,Andrade:2018eta,Zhao:2013yaa}. Another common method is the dipole fitting method~\cite{Mariano:2012wx}, which assumes a priori \Mov{the existence of a }dipole in the cosmic anisotropy on small scales~\cite{Deng:2018yhb,Sun:2018cha,Deng:2018jrp,Lin:2015rza,Zhao:2019azy}. Until now, there is no evidence for cosmic anisotropy on small scales in the SNIa samples. However, these null signals are just \Mov{obtained from }overall constraints and may neglect some fine structures in the universe.

\begin{figure*}[]
\begin{center}
\subfloat{\includegraphics[width=0.5\textwidth]{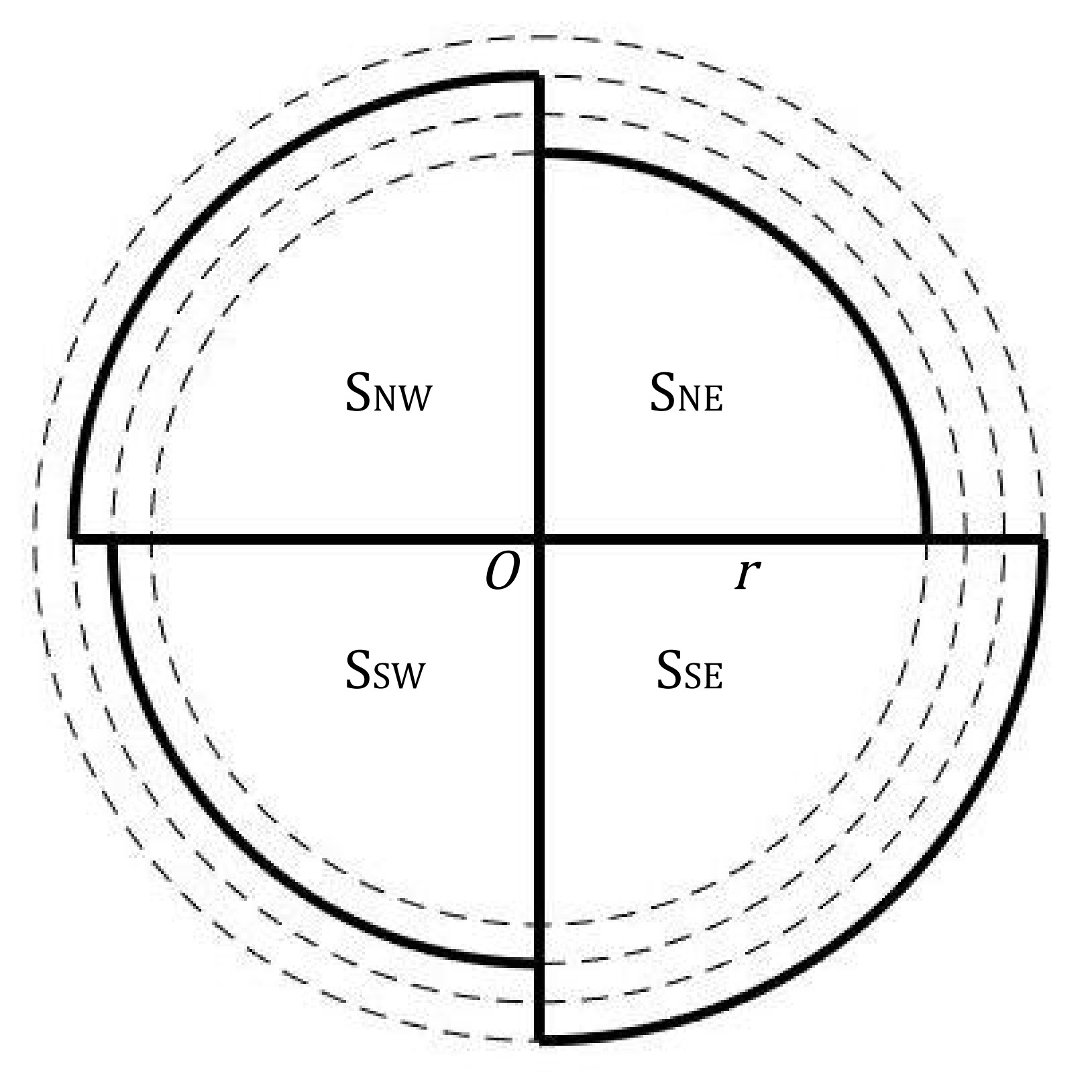}}
\end{center}
\captionsetup{justification=raggedright}
\caption{The diagram of an irregular local cosmic void. Each segment with the same solid angle $\pi$  is modeled by its own LTB metric and constrained by the corresponding SNIa subset given in Tab.~\ref{tb:sample}. \MovB{Different specific solid angle LTB segments calibrated on data from that angle allow to capture anisotropic fine structures, contrary to previous isotropic studies \cite{Camarena:2021mjr,Camarena:2022iae}. Those studies' spherical
symmetry assumption restriction might have washed away a weak signal from some fine structures. Such signal would favor the $\Lambda$LTB model over the $\Lambda$CDM model.}}
\label{fig:void}
\end{figure*}

In this paper, we try to test both of the local cosmic inhomogeneity and the cosmic anisotropy on small scales at the same time using the SNIa data of Pantheon. To account for the asymmetry in the local cosmic inhomogeneity and the fine structures in the cosmic anisotropy on small scales, we will \Mov{fit }cosmic anisotropy on small scales with the $\Lambda$LTB model. 
\Mov{We first divide} the local cosmic void into several segments. Each segment with \Mov{given} solid angle is \Mov{fitted to} its own LTB metric{, where the \MovE{void depth and radius }parameters only \MovE{correspond to }the local segment and \MovE{ignore the \MovB{data} from} the other segments}. \Mov{We then} divide the 1048 SNIa of Pantheon into corresponding subsets according to their distribution in the galactic coordinate system. Obviously, each SNIa subset can only be used to reconstruct the profile of one segment. Finally, we can patch together an irregular profile for the local cosmic void with the whole Pantheon sample. The whole profile will contain all the information about both of the local cosmic inhomogeneity and the cosmic anisotropy on small scales, as shown in Fig.~\ref{fig:void}.

This paper is organized as follows.
In section~\ref{sec:MD}, we present our method modeling the local cosmic inhomogeneity and introduce our treatment of the Pantheon data.
In section~\ref{sec:R}, we show the constraints on the profiles of all segments of the local cosmic inhomogeneity and compare them.
Finally, a brief summary and discussions are included in section~\ref{sec:SD}.

\section{Methodology and Data}
\label{sec:MD}
\subsection{Model}
\label{ss:M}
A spherically symmetric void can be modeled by the LTB metric 
\begin{equation}
ds^2=-dt^2+\frac{R'^2(r,t)}{1+2r^2k(r)\tilde{M}^2}dr^2+R^2(r,t)d\Omega,
\end{equation}
where $d\Omega = d\theta^2+\sin^2\theta d\phi^2$, $\tilde{M}$ is an arbitrary mass scale, $k(r)$ is an arbitrary curvature profile function, $R(r,t)=a(t)r$ is dependent on the FLRW scale factor $a(t)$ and a prime (or dot) denotes derivative with respect to the radial coordinate $r$ (or the time $t$). As $\Lambda$CDM model is built on the FLRW metric, there is a $\Lambda$LTB model built on the LTB metric. The universe's expansion in the $\Lambda$LTB model is determined by a Friedmann-like equation
\begin{equation}
\frac{\dot{R}^2(r,t)}{R^2(r,t)}=\frac{2m(r)}{R^3(r,t)}+\frac{2r^2k(r)\tilde{M}^2}{R^2(r,t)}+\frac{\Lambda}{3},
\end{equation}
where $m(r)$ is the so-called Euclidean mass function\footnote{\Mov{Choice of initial density profile.}} and be set as $m(r) = 4\pi \tilde{M}^2r^3/3$, the Big Bang time $t_{\rm BB}(r)$ is fixed as a constant and the curvature profile $k(r)$ is the only free function to determine the void. More precisely, \Mov{the void's expansion can be decomposed into} a transverse expansion rate and a longitudinal expansion rate. The former one \Mov{depends} on a transverse scale factor $a_{\bot}(r,t)=R(r,t)/r$ as
\begin{align}
H_{\bot}(r,t)=&\frac{\dot{a}_{\bot}(r,t)}{a_{\bot}(r,t)};
\end{align}
\Mov{while }the latter one is \Mov{defined by} a longitudinal scale factor $a_{\parallel}(r,t)=R'(r,t)$ as
\begin{align}
H_{\parallel}(r,t)=&\frac{\dot{a}_{\parallel}(r,t)}{a_{\parallel}(r,t)}.
\end{align}
According to the above Friedmann-like equation, we can define the density parameters of matter, curvature and dark energy today as
\begin{align}
\Omega_{\rm m}(r)=&\frac{2m(r)}{R^3(r,t_0)H_{\bot}^2(r,t_0)},\\
\Omega_{\rm k}(r)=&\frac{2r^2k(r)\tilde{M}^2}{R^2(r,t_0)H_{\bot}^2(r,t_0)},\\
\Omega_{\rm \Lambda}(r)=&\frac{\Lambda}{3H_{\bot}^2(r,t_0)}.
\end{align}

The profile of the void can be parameterized by the depth\Mov{,} size\Mov{,} and \Mov{boundary }width of the void~\cite{Garcia-Bellido:2008vdn}. In fact, the depth and size of the void are \Mov{sufficient to constrain it } and the width of the void boundary can be ignored~\cite{Valkenburg:2012td}\Mov{. We thus parameterise the void with the following curvature profile}
\begin{align}
k(r)&=k_{\rm c}P_3(r),\\
P_n(r)&=\left \{
\begin{array}{lcl}
1-\exp\left[\frac{-(1-r/r_{\rm B})^n}{r/r_{\rm B}}\right] && 0 \leq r < r_{\rm B} \\
0 && r_{\rm B} \leq r \\
\end{array} \right.,\label{eq:ShapeFunction}
\end{align}
where we have assumed the universe is flat outside the void, $k_{\rm c}$ is the curvature at the center and $r_{\rm B}$ is the comoving radius of the void.

If the void is spherically symmetric and \Mov{all the} latest 
cosmological observations are \Mov{available}, the profile at $r$ can be constrained by data from any direction or sky location. Therefore, the depth of the void and the boundary of the void in particular\Mov{,} where the curvature \Mov{reaches} $0$\Mov{,} can be constrained relatively well~\cite{Camarena:2021mjr}.
If the asymmetry of the void\Mov{ is only probed with SNIa data}, however, we find \Mov{that although }the depth of the void \Mov{can still } be constrained relatively well\Mov{, this is not the case for} the boundary of the void\Mov{,} where the curvature changes to $0$.
Therefore, we introduce a \Mov{scale }$r_{\rm C}$ where the curvature changes by $10\%$. That is to say, we will not attempt to \Mov{characterise} the whole void but conservatively concentrate on the partial profile of the \Mov{inner }void\Mov{, replacing Eq.~\eqref{eq:ShapeFunction} with} the following function:
\begin{equation}
P_{ n}(r) =\left \{
\begin{array}{lcl}
1-0.1\exp \left[ \frac{-\left(1-r/r_{\rm C}\right)^n}{r/r_{\rm C}}\right] && 0\leq r <r_{\rm C} \\
0.9\left[1-\exp\left[\frac{-\left(1-\frac{r-r_{\rm C}}{r_{\rm B}-r_{\rm C}}\right)^n}{\frac{r-r_{\rm C}}{r_{\rm B}-r_{\rm C}}}\right]\right] && r_{\rm C}\leq r <r_{\rm B} \\
0 && r_{\rm B} \leq r \\
\end{array} \right..
\end{equation}

\Mov{The} void \Mov{now }is parameterized by three parameters $\{k_{\rm c},r_{\rm C},r_{\rm B}$\}. All of them are derived parameters in our code. $k_{\rm c}$ is related to $\tilde{\delta}_0=\tilde{\delta}(r=0,t_0)$, where the integrated mass density contrast $\tilde{\delta}(r,t_0)$~\cite{Camarena:2021mjr} is defined as 
\begin{equation}
\tilde{\delta}(r,t_0)= \frac{\Omega_m{H_{\bot}}^2}{\Omega_m^{\rm out}{H_0^{\rm out}}^2}-1,
\end{equation}
where ``out" denotes the corresponding FLRW quantities. Because $-1\leq \tilde{\delta}_0<\infty$ is not good for the convergence of the Monte Carlo Markov Chain (MCMC), we will use a new parameter
\begin{equation}
\delta_0 =\left \{
\begin{array}{lcl}
\tilde{\delta}_0 && \tilde{\delta}_0 \leq 0 \\
\tilde{\delta}_0/(1+\tilde{\delta}_0) && \tilde{\delta}_0 >0 \\
\end{array} \right..
\end{equation}
As for the two radii $\{r_{\rm B},r_{\rm C}\}$, we will relate them to their corresponding redshifts by $r_{\rm B}=r(z_{\rm B})$ and $r_{\rm C}=r(z_{\rm C})$, where $r(z)$ satisfies the geodesic equations
\begin{align}
\frac{dt}{dz}=&-\frac{R'(r,t)}{(1+z)\dot{R'}(r,t)},\\
\frac{dr}{dz}=&-\frac{\sqrt{1+2r^2k(r)\tilde{M}^2}}{(1+z)\dot{R'}(r,t)}.
\end{align}
As mentioned before, the information about the whole void is out of reach of SNIa data. \MovE{As we probe the central part of the void with an evaluated $r_{\rm C}$ and consider the unknown boundary to be far, but not extremely far, from that limit, we therefore} further relate $r_{\rm B}$ to $r_{\rm C}$ as $r_{\rm B}=2r_{\rm C}$. That is to say, we will only conservatively concentrate on the profile of the void \Mov{where} $z\leq z_{\rm C}$\Mov{, while} the profile of the void \Mov{where} $z_{\rm C}<z\leq z_{\rm B}$ is set by our assumption $r_{\rm B}=2r_{\rm C}$. Moreover, \Mov{as }$z_{\rm B}>0$ is meaningless when $\delta_0\sim0$, we relate $z_{\rm B}$ to $|\delta_0|$ by a free parameter $|z_{\rm B}/\delta_0|$ which will \Mov{proceed from a} uniform prior distribution $|z_{\rm B}/\delta_0|\in[0,100]$. The \Mov{remaining} free parameter $\delta_0$ will \Mov{have a} uniform prior distribution $\delta_0\in[-0.99,0.99]$. \Mov{We finally} can use the dependence of SNIa's luminosity distance on these two \Mov{synthetic }void parameters $\{\delta_{0},z_{\rm B}=|\delta_{0}|\times|z_{\rm B}/\delta_0|\}$ to probe the profile of the void, where the angular diameter distance $d_{\rm A}$ and the luminosity distance $d_{\rm L}$ are obtained\Mov{ from}
\begin{align}
d_{\rm A}(z;\delta_{0},z_{\rm B})=&R(r(z),t(z);\delta_{0},z_{\rm B}), \\
d_{\rm L}(z;\delta_{0},z_{\rm B})=&(1+z)^2 d_{\rm A}(z;\delta_{0},z_{\rm B}).
\end{align}.

\subsection{Data}
\label{ss:D}
\begin{figure*}[]
\begin{center}
\subfloat{\includegraphics[width=0.95\textwidth]{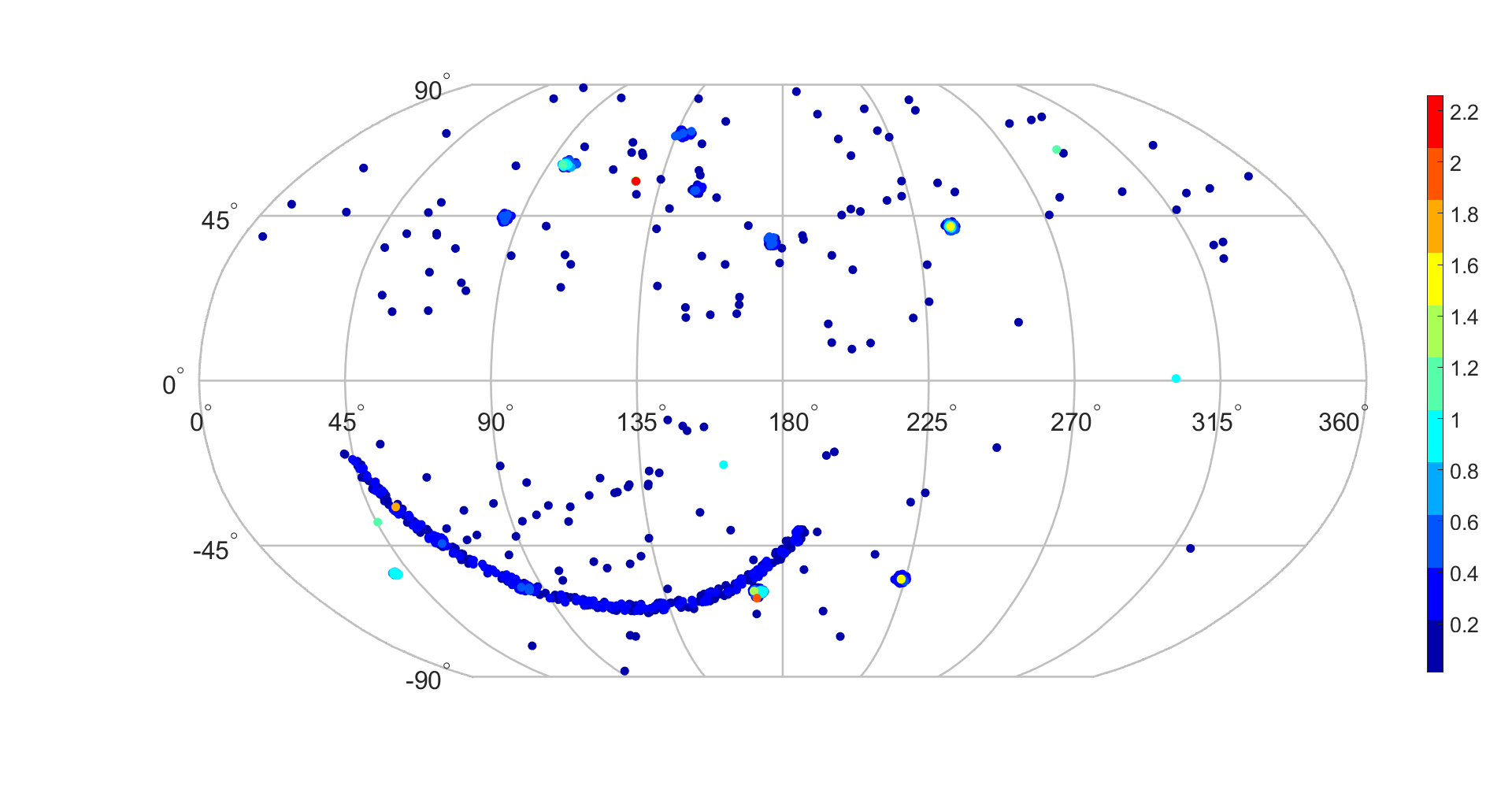}}
\end{center}
\captionsetup{justification=raggedright}
\caption{The distribution of 1048 SNIa of Pantheon in the galactic coordinate system $(l,b)$, where the discrete colorbar indicates the redshifts of these SNIa.}
\label{fig:sample}
\end{figure*}

In this paper, we only use the combined Pantheon sample~\cite{Pan-STARRS1:2017jku} to constrain the local cosmic void. This dataset consists of 1048 SNIa in the redshift range $0.01<z<2.3$. In Fig.~\ref{fig:sample}, we show the distribution of these 1048 SNIa in the galactic coordinate system $(l,b)$. To probe the asymmetry in the local cosmic inhomogeneity, we divide the full dataset into several subsets, as listed in Tab.~\ref{tb:sample}. Obviously, each SNIa subset can only give the local information of the universe which can be characterized by a corresponding LTB metric. \Mov{That} is to say, the whole profile of the local comic void should be reconstructed with the full SNIa dataset and described by several corresponding LTB metrics. \MovB{In other words, our model obtains a probe of anisotropy as illustrated in Fig.~\ref{fig:void}. There, each LTB metric being spherically symmetric, the anisotropy of our model does not proceed from individual segments but from the assembly of the various solid angle segments. The model for each segment of the final assembly is indeed spherically symmetric. Such segment’s model is obtained by considering the data from that specific solid angle to virtually be duplicated in all directions into a spherically symmetric virtual data coverage. It then can be represented by an LTB metric. We virtually proceed by restricting the obtained LTB metric within its source data solid angle. The set of such solid-angle-restricted LTB metric is then able to capture anisotropy in the composite total model, despite originating from spherically symmetric models. }

Therefore, for \Mov{the case when the whole data set is used (hereafter, }no cut case\Mov{)}, we have
\begin{equation}
\chi^2 =\chi^2_{\rm P_{\rm All}};
\end{equation}
for \Mov{the division between North and South galactic plane subsets (}one horizontal cut case\Mov{)}, we have 
\begin{equation}
\chi^2=\chi^2_{\rm P_{\rm N}}+\chi^2_{\rm P_{\rm S}};
\end{equation}
for \Mov{the division between East and West galactic plane subsets (}one vertical cut case\Mov{)}, we have 
\begin{equation}
\chi^2=\chi^2_{\rm P_{\rm E}}+\chi^2_{\rm P_{\rm W}};
\end{equation}
for \Mov{the division into four quadrant using the previous subsets of the galactic plane (}two cuts case\Mov{)}, we have 
\begin{equation}
\chi^2 =\chi^2_{\rm P_{\rm NE}}+\chi^2_{\rm P_{\rm NW}}+\chi^2_{\rm P_{\rm SE}}+\chi^2_{\rm P_{\rm SW}}.
\end{equation}
The $\chi^2_s$ for every data subset \Mov{are defined with}
\begin{equation}
\chi^2_s=\sum(\bm{m}_{{\rm B},s}^{\rm obs}-\bm{m}_{{\rm B},s}^{\rm mod})^T\bm{C}^{-1}_s(\bm{m}_{{\rm B},s}^{\rm obs}-\bm{m}_{{\rm B},s}^{\rm mod}),
\end{equation}
where $\bm{C}_s$ is the covariance matrix\MovB{\footnote{\MovB{The covariant matrix is made of all SNIa data correlations with each other for the total set. Therefore the subset correlation matrices are simply the sub-blocks of the total matrix where the lines and columns have been rearranged to group the selected subsets together, restricted to the subset considered.}}} of the $s$-th subset, the apparent magnitude\Mov{s} $\bm{m}_{{\rm B},s}^{\rm obs}$ observed by Pantheon\Mov{, including contributions from} stretch $x_1$, color $c$ and host-galaxy correction $\Delta M$\Mov{. Note that} the apparent magnitude $m_{{\rm B},s,i}^{\rm mod}$ for the $i$-th SNIa of the $s$-th subset in the $\Lambda$LTB model \Mov{depends} on two void parameters $\{\delta_{0,s},z_{{\rm B},s}\}$ and the absolute magnitude $M_{\rm B}$ as
\begin{multline}
m_{{\rm B},s,i}^{{\rm mod}}(z_{i};\delta_{0,s},z_{{\rm B},s},M_{{\rm B}})\\
=5\log_{10}\frac{d_{{\rm L}}(z_{i};\delta_{0,s},z_{{\rm B},s})}{1{\rm Mpc}}+25+M_{{\rm B}}.
\end{multline}
Since SNIa are supposed to be standard candles, we exclude the effects of variations of $M_{\rm B}$ in each SNIa subset on the asymmetry in the local cosmic inhomogeneity. Therefore, for every \Mov{division} case, we use the full dataset to constrain the only nuisance parameter $M_{\rm B}$. 

\begin{table}
\captionsetup{justification=raggedright}
\caption{According to the distribution of 1048 SNIa of Pantheon~\cite{Pan-STARRS1:2017jku} in the galactic coordinate system $(l,b)$, the full dataset is divide into $ \rm P_{\rm N}$+$ \rm P_{\rm S}$ by \Mov{a division along the galactic plane}, $ \rm P_{\rm E}$+$ \rm P_{\rm W}$ by \Mov{a division orthogonal to the galactic plane} and $ \rm P_{\rm NE}$+$ \rm P_{\rm NW}$+$ \rm P_{\rm SE}$+$ \rm P_{\rm SW}$ by \Mov{both simultaneous divisions}.}
\label{tb:sample}
\begin{tabular}{ccc}
\hline
\hline
\rm Dataset & $l$ &  $b$ \\
\hline
$ \rm P_{\rm All}$ &  0°$\leq l < $360° &  $-90$°$\leq b \leq $90°  \\
\hline
$ \rm P_{\rm N}$ &  0°$\leq l < $360° &  0°$\leq b \leq $90°   \\
$ \rm P_{\rm S}$ &  0°$\leq l < $360° &  $-90$°$\leq b <$0°   \\
\hline
$ \rm P_{\rm W}$  &  0°$\leq l \leq $180°&  $-90$°$\leq b \leq $90°   \\
$ \rm P_{\rm E}$  &  180°$< l < $360°  &  $-90$°$\leq b \leq $90°  \\
\hline
$ \rm P_{\rm NW}$  &  0°$\leq l \leq $180° &  0°$\leq b \leq $90°   \\
$ \rm P_{\rm NE}$  &  180°$< l < $360° &  0°$\leq b \leq $90°   \\
$ \rm P_{\rm SW}$  &  0°$\leq l \leq $180° &  $-90$°$\leq b <$0°   \\
$ \rm P_{\rm SE}$  &  180°$< l < $360° &  $-90$°$\leq b <$0°   \\
\hline 
\hline
\end{tabular}
\end{table}

\section{Results}
\label{sec:R}
The luminosity distance $d_{\rm L}(z;\delta_{0},z_{\rm B})$ in the $\Lambda$LTB model is numerically given by $\mathsf{VoidDistances2020}$~\cite{Valkenburg:2011tm} which should be fed with parameters \Mov{corresponding to the largest scales of the universe}. The initialization of the \Mov{large scale }universe \Mov{parameters, }outside the void\Mov{,} is done by $\mathsf{CLASS}$~\cite{Blas:2011rf}, even though we don't use the CMB data here. For $\mathsf{CLASS}$~\cite{Blas:2011rf}, we need to provide the $\Lambda$CDM model's six parameters. In Tab.~\ref{tb:bk}, we list \Mov{such} parameters given by Planck 2018 \Mov{for }TT,TE,EE+lowE+lensing~\cite{Planck:2018vyg}. Finally, the likelihoods in subsection~\ref{ss:D} are added into $\mathsf{MontePython}$~\cite{Audren:2012wb} by our modified $\mathsf{monteLLTB}$~\cite{Camarena:2021mjr}.
\begin{table}
\captionsetup{justification=raggedright}
\caption{The $\Lambda$CDM model's six parameters given by Planck 2018 TT,TE,EE+lowE+lensing~\cite{Planck:2018vyg}.}
\label{tb:bk}
\begin{tabular}{cccccc}
\hline
\hline
$\Omega_{\rm b}h^2$ & $\Omega_{\rm c}h^2$ &  $H_0[{\rm km~s^{-1}~Mpc^{-1}}]$  &  $\tau$  &   $\ln10^{10}A_{\rm s}$  &  $n_{\rm s}$ \\
\hline 
$0.02237$  &  0.12  & 67.36   &  0.0544  &  3.044  &  0.9649  \\
\hline
\hline
\end{tabular}
\end{table}

\begin{figure*}[]
\begin{center}
\subfloat{\includegraphics[width=.95\textwidth]{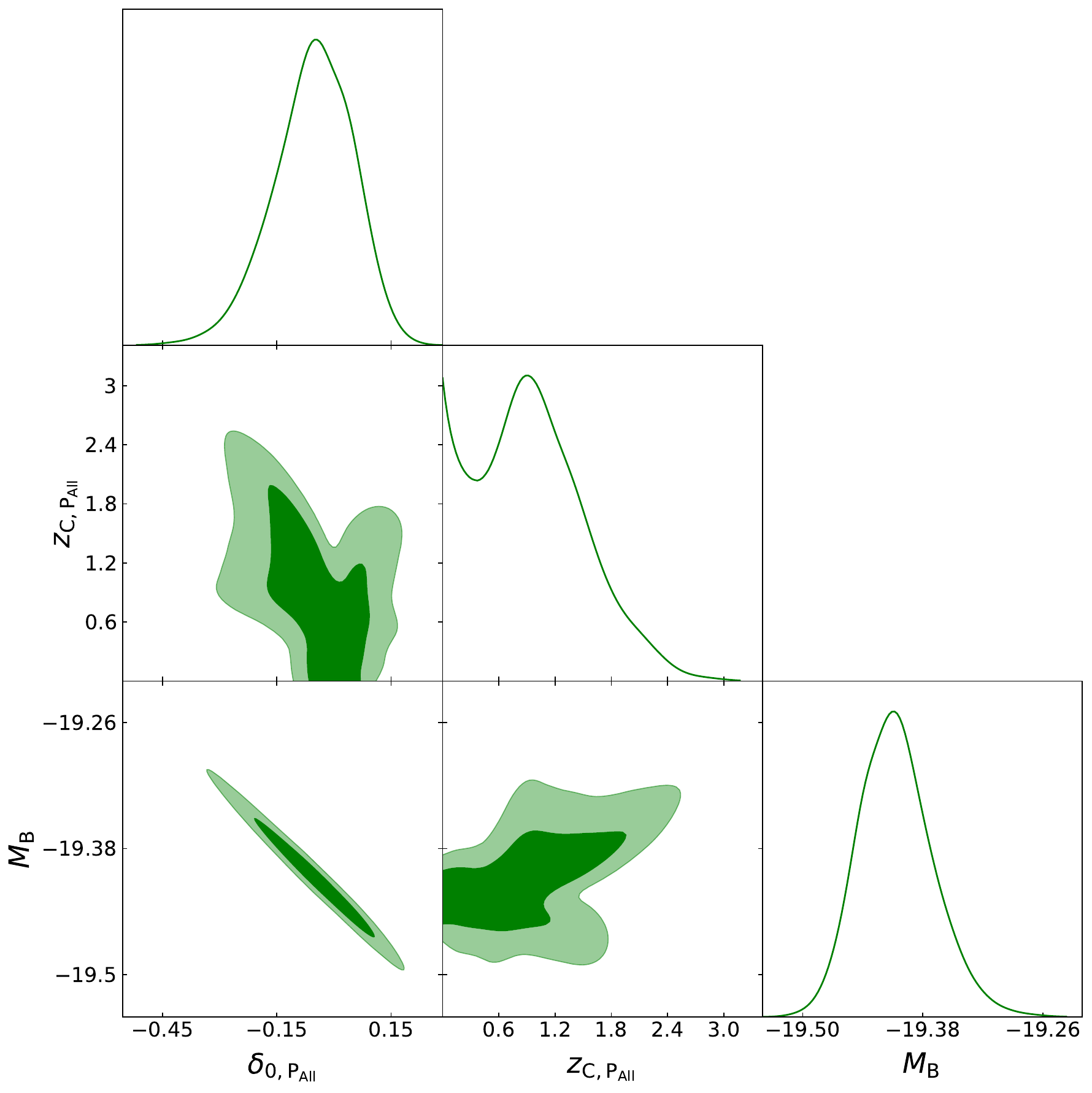} }
\end{center}
\captionsetup{justification=raggedright}
\caption{The triangular plot of void parameters\MovE{ (see Sec.~\ref{sec:R})} and $M_{\rm B}$ for \Mov{the }no cut case, where the contours \MovE{shown at $68\%$ (inner lines) and $95\%$ (outer lines)} confidence ranges.}
\label{fig:pall}
\end{figure*}

\begin{figure*}[]
\begin{center}
\subfloat{\includegraphics[width=.95\textwidth]{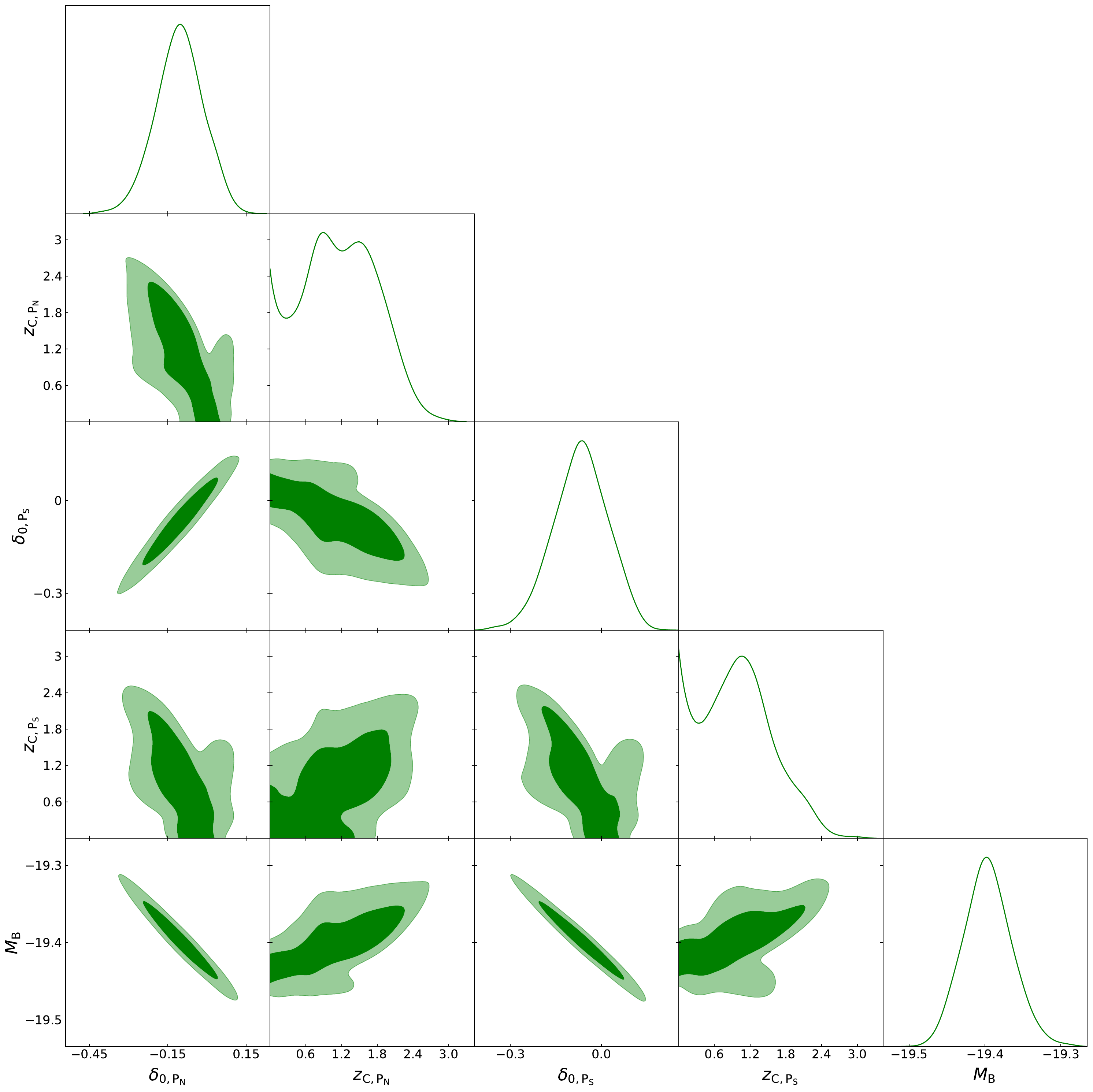} }
\end{center}
\captionsetup{justification=raggedright}
\caption{The triangular plot of void parameters\MovE{ (see Sec.~\ref{sec:R})} and $M_{\rm B}$ for \Mov{the} horizontal cut case, where the contours \MovE{shown at $68\%$ (inner lines) and $95\%$ (outer lines)} confidence ranges.}
\label{fig:pns}
\end{figure*}

\begin{figure*}[]
\begin{center}
\subfloat{\includegraphics[width=.95\textwidth]{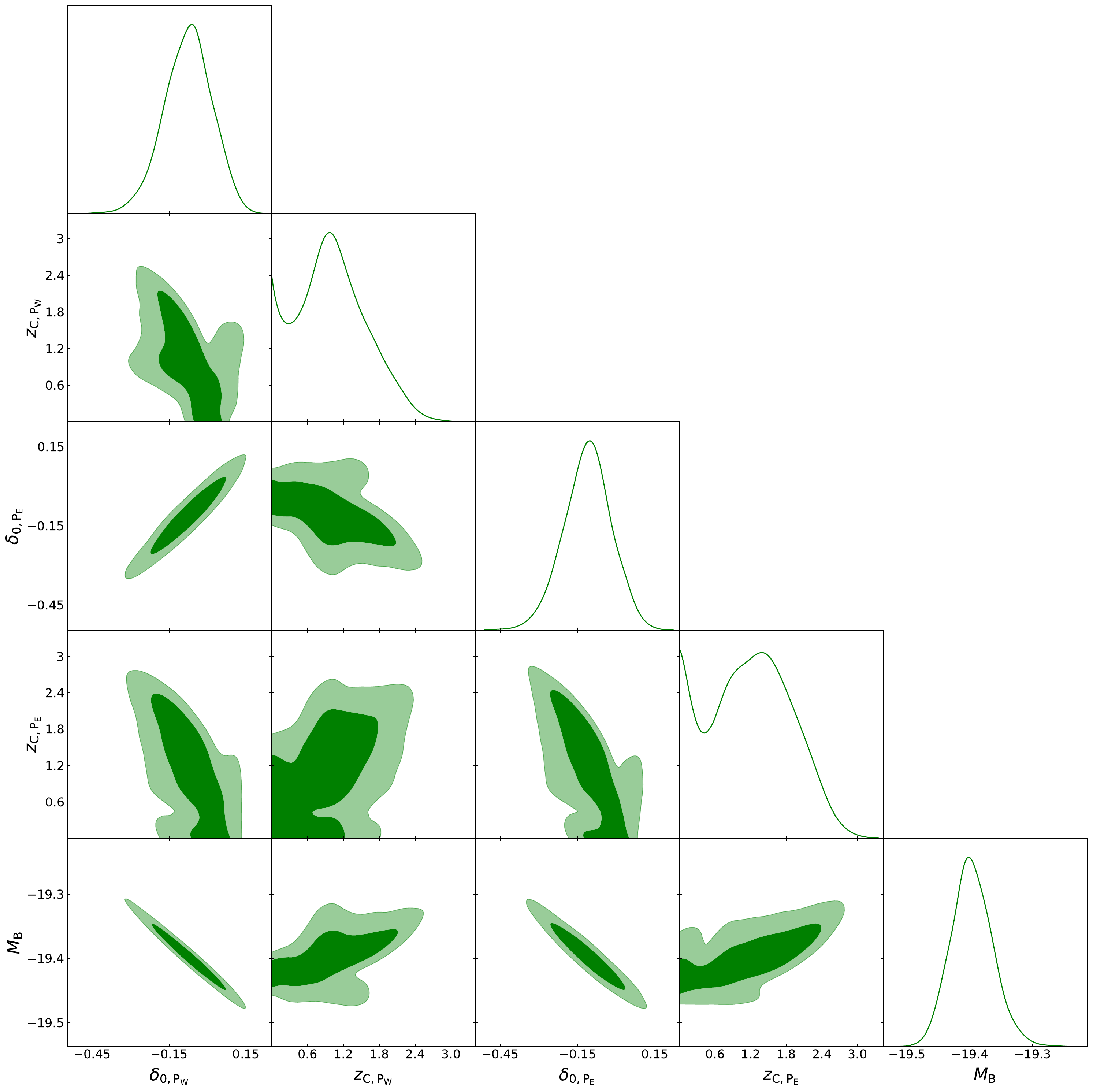} }
\end{center}
\captionsetup{justification=raggedright}
\caption{The triangular plot of void parameters\MovE{ (see Sec.~\ref{sec:R})} and $M_{\rm B}$ for \Mov{the} vertical cut case, where the contours \MovE{shown at $68\%$ (inner lines) and $95\%$ (outer lines)} confidence ranges.}
\label{fig:pwe}
\end{figure*}

\begin{figure*}[]
\begin{center}
\subfloat{\includegraphics[width=.95\textwidth]{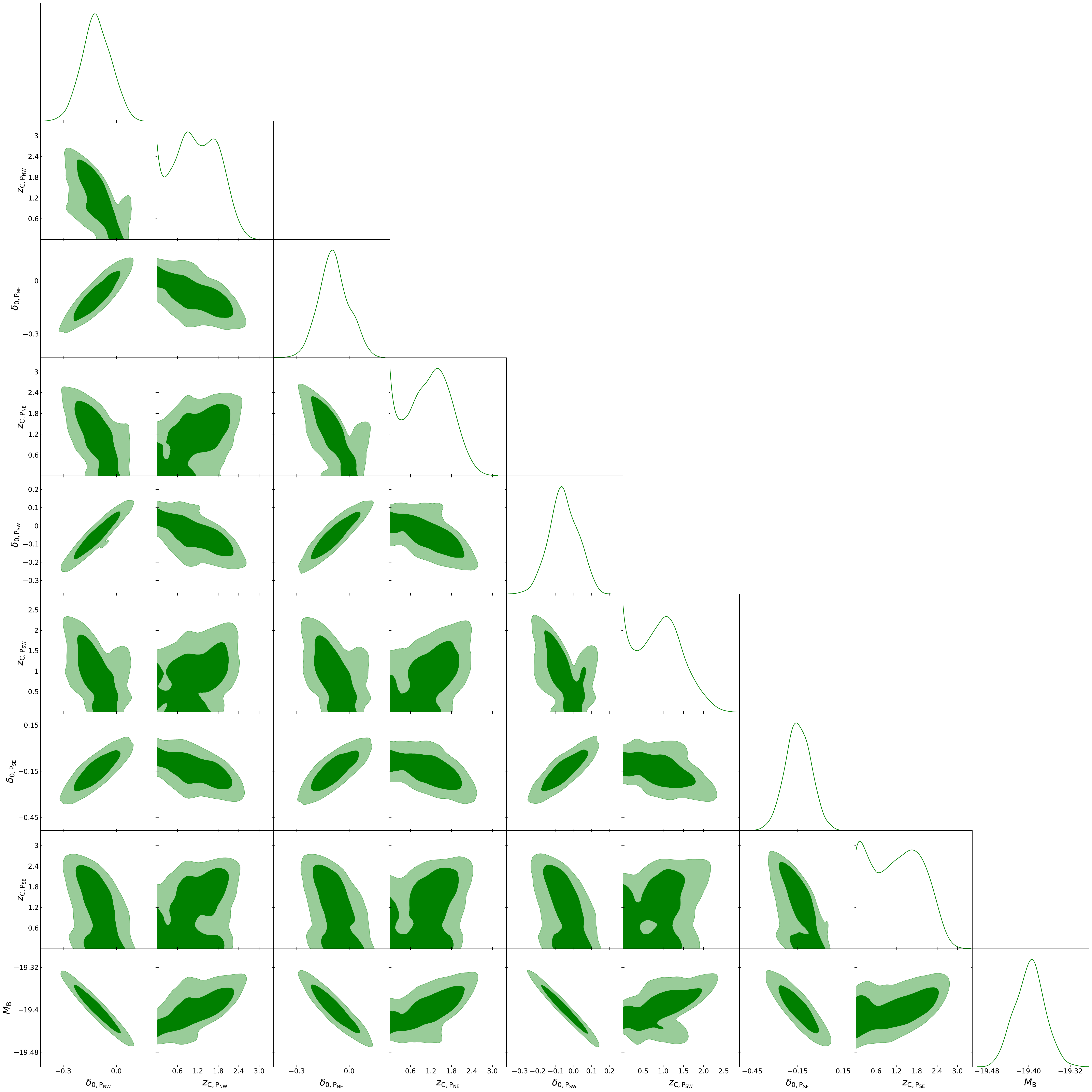} }
\end{center}
\captionsetup{justification=raggedright}
\caption{The triangular plot of void parameters\MovE{ (see Sec.~\ref{sec:R})} and $M_{\rm B}$ for \Mov{the }two cuts case, where the contours \MovE{shown at $68\%$ (inner lines) and $95\%$ (outer lines)} confidence ranges.}
\label{fig:pnswe}
\end{figure*}

In Tab.~\ref{tb:result}, the constraints on $\{\delta_{0,s},z_{{\rm C},s},M_{\rm B}\}$ with every data subset are summarized.
In Fig\Mov{s}.~\ref{fig:pall}, \ref{fig:pns}, \ref{fig:pwe} and \ref{fig:pnswe}, the constraints on $\{\delta_{0,s},z_{{\rm C},s},M_{\rm B}\}$ with every data subset for \Mov{the }no cut, horizontal cut, vertical cut and two cuts case\Mov{s} are also shown respectively. \MovE{We choose to assume no anisotropy for $M_B$ and treat it as a nuisance parameter. Because we need a standard probe, we used the full data set to constrain it. This is vindicated  as the results shown in Tab.~\ref{tb:result} turn out to be  self-consistent.}
Although we suppose $\delta_{0,s}$ of every LTB metric is only constrained with the corresponding data subset and there is no correlation between $\delta_{0,s}$, for any \Mov{division} case, there is an obvious correlation between $\delta_{0,s}$ in Fig\Mov{s}.~\ref{fig:pns}, \ref{fig:pwe} and \ref{fig:pnswe}. That \Mov{correlation results from our assumption that} SNIa are standard candles \Mov{so} all $\delta_{0,s}$ directly correlates with $M_{\rm B}$ and \Mov{thus }indirectly correlates with the other $\delta_{0,s}$. For every \Mov{division} case, the correlation between $\delta_{0,s}$ just leads to a similar error \Mov{on} $\delta_{0,s}$ $0.08\sim0.10$ but imposes no effect on the mean value of $\delta_{0,s}$. Generally speaking, for all \Mov{division} cases, the constraints on $\delta_{0,s}$ are consistent with the FLRW metric at $95\%$ confidence level (CL): on the one hand, the constraints on $\delta_{0,s}$ are consistent with 0, which \Mov{denotes} cosmic homogeneity; on the other hand, the constraints on $\delta_{0,s}$ are consistent with each other, which \Mov{indicates} cosmic isotropy. However, the constraint on $\delta_{\rm 0,P_{SE}}=-0.15\pm0.08$ deviates from 0 at almost $2\sigma$. This deviation results \Mov{either }from the paucity of data in \Mov{the }${\rm P_{SE}}$ data subset or the real depth of the local cosmic void in this direction. Even though we have given up \Mov{the determination of }the profile at $z_{\rm C}\leq z<z_{\rm B}$ by setting $r_{\rm B}=2r_{\rm C}$, the constraints on $z_{{\rm C},s}$ are still very weak. \Mov{Although we found PDF peaks at $z_{{\rm C},s}=0$ and $1\lesssim z_{{\rm C},s}\lesssim 2.26$,} we can't conclude that the Pantheon data prefers a non-zero $z_{{\rm C},s}$ to $z_{{\rm C},s}=0$.

\begin{figure*}[]
\begin{center}
\subfloat{\includegraphics[width=.95\textwidth]{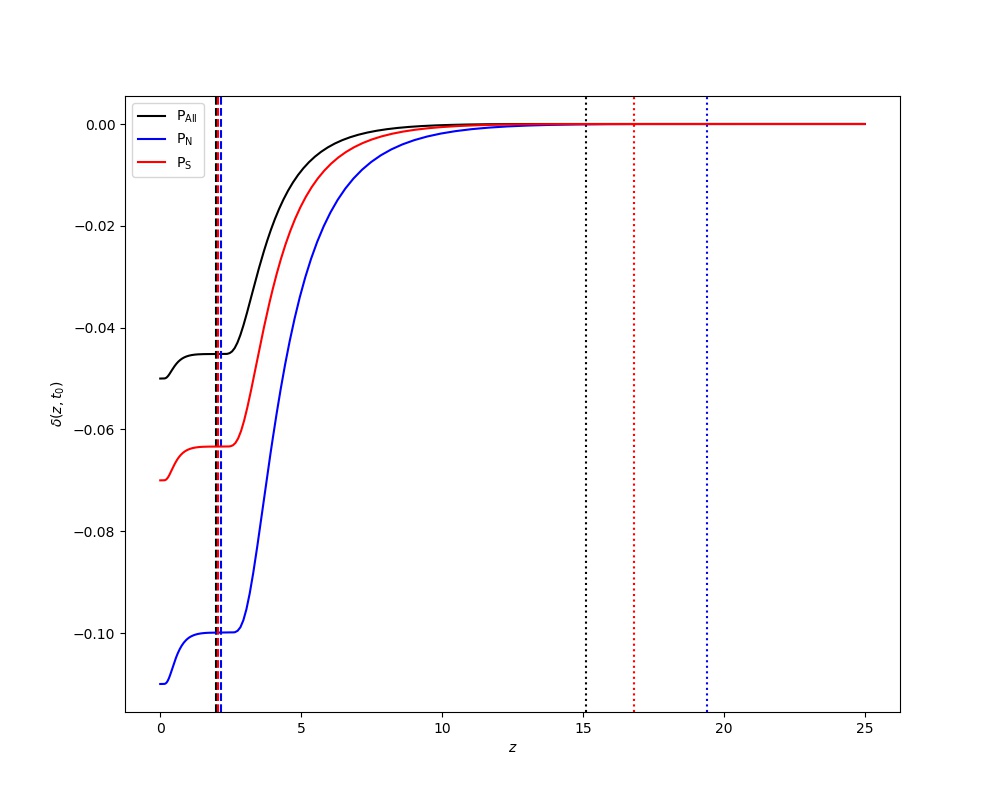} }
\end{center}
\captionsetup{justification=raggedright}
\caption{The integrated mass density contrast $\delta(z,t_0)$ for \Mov{the }no cut and horizontal cut case\Mov{s}, where $\delta_0=-0.05$ (black solid), $z_{\rm C}= 1.97$ (black dashed) and $z_{\rm B}= 15.1$ (black dotted) for $ \rm P_{\rm All}$, $\delta_0=-0.11$ (blue solid), $z_{\rm C}= 2.16$ (blue dashed) and $z_{\rm B}= 19.4$ (blue dotted) for $ \rm P_{\rm N}$ and $\delta_0=-0.07$ (red solid), $z_{\rm C}= 2.05$ (red dashed) and $z_{\rm B}= 16.8$ (red dotted) for $ \rm P_{\rm S}$ respectively.}
\label{fig:pallns}
\end{figure*}

\begin{figure*}[]
\begin{center}
\subfloat{\includegraphics[width=.95\textwidth]{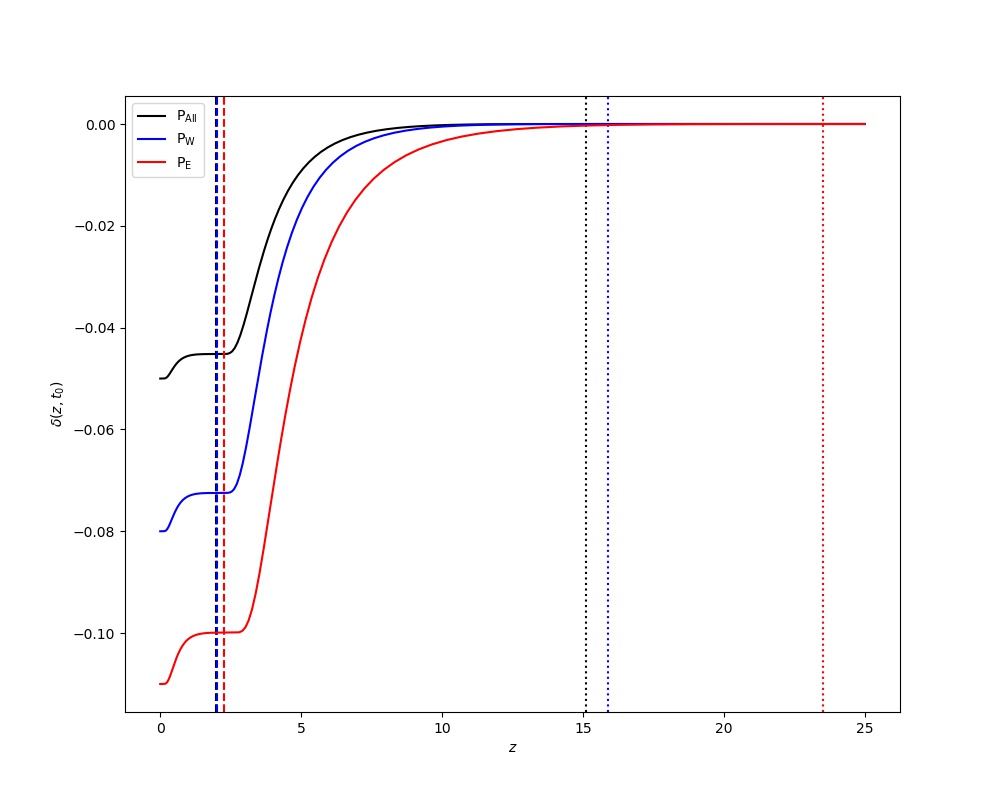} }
\end{center}
\captionsetup{justification=raggedright}
\caption{The integrated mass density contrast $\delta(z,t_0)$ for \Mov{the }no cut and vertical cut case\Mov{s}, where $\delta_0=-0.05$ (black solid), $z_{\rm C}= 1.97$ (black dashed) and $z_{\rm B}= 15.1$ (black dotted) for $ \rm P_{\rm All}$, $\delta_0=-0.08$ (blue solid), $z_{\rm C}= 2.02$ (blue dashed) and $z_{\rm B}= 15.9$ (blue dotted) for $ \rm P_{\rm W}$ and $\delta_0=-0.11$ (red solid), $z_{\rm C}= 2.28$ (red dashed) and $z_{\rm B}= 23.5$ (red dotted) for $ \rm P_{\rm E}$ respectively.}
\label{fig:pallwe}
\end{figure*}

\begin{figure*}[]
\begin{center}
\subfloat{\includegraphics[width=.95\textwidth]{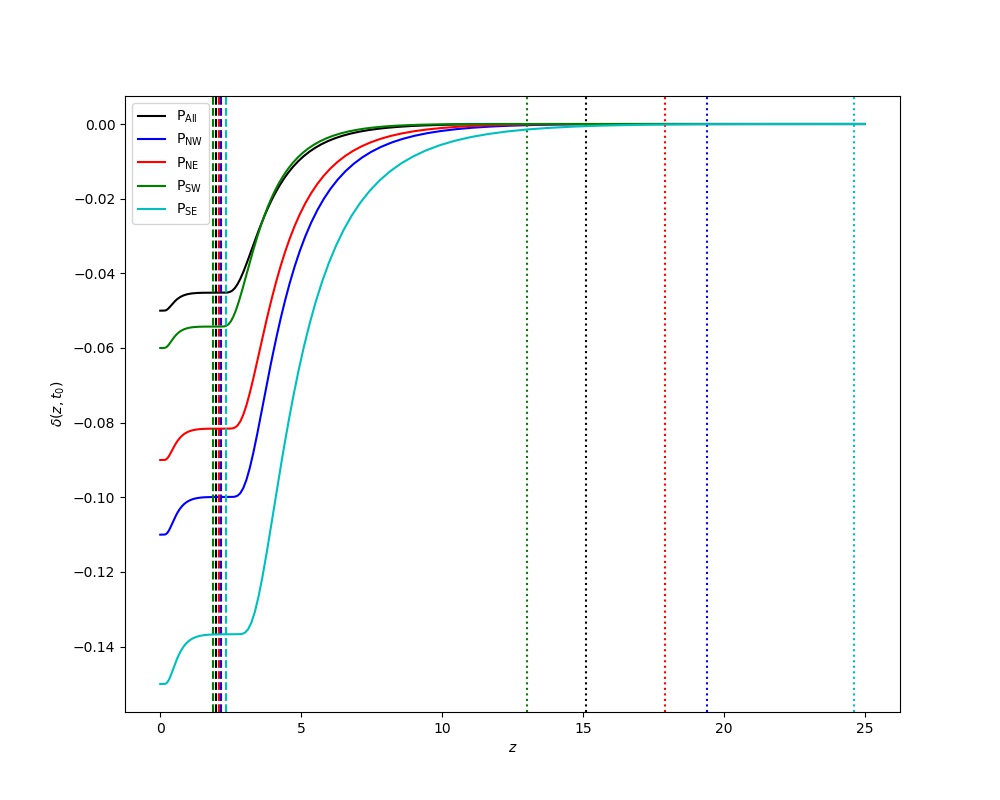} }
\end{center}
\captionsetup{justification=raggedright}
\caption{The integrated mass density contrast $\delta(z,t_0)$ for \Mov{the }no cut and two cuts case\Mov{s}, where $\delta_0=-0.05$ (black solid), $z_{\rm C}= 1.97$ (black dashed) and $z_{\rm B}= 15.1$ (black dotted) for $ \rm P_{\rm All}$, $\delta_0=-0.11$ (blue solid), $z_{\rm C}= 2.16$ (blue dashed) and $z_{\rm B}= 19.4$ (blue dotted) for $ \rm P_{\rm NW}$, $\delta_0=-0.09$ (red solid), $z_{\rm C}= 2.10$ (red dashed) and $z_{\rm B}= 17.9$ (red dotted) for $ \rm P_{\rm NE}$, $\delta_0=-0.06$ (green solid), $z_{\rm C}= 1.88$ (green dashed) and $z_{\rm B}= 13.0$ (green dotted) for $ \rm P_{\rm SW}$ and $\delta_0=-0.15$ (cyan solid), $z_{\rm C}= 2.34$ (cyan dashed) and $z_{\rm B}= 24.6$ (cyan dotted) for $ \rm P_{\rm SE}$ respectively.}
\label{fig:pallnswe}
\end{figure*}

Finally, we can use the constraints on $\{\delta_{0,s},z_{{\rm C},s}\}$ for every \Mov{division} case\MovE{, i.e. their best fit,} to probe both of the local cosmic inhomogeneity and the cosmic anisotropy on small scales. In Fig\Mov{s}.~\ref{fig:pallns}, \ref{fig:pallwe} and \ref{fig:pallnswe}, we reconstruct the profile of local cosmic void $\delta(z,t_0)$ in different direction (solid lines) with the constraints summarized in Tab.~\ref{tb:result}. At $z\lesssim z_{{\rm C},s}$ (dashed lines), the curvature changes by $10\%$. And we complete the rest of profile until $z_{{\rm B},s}$ (dotted lines) by setting $r_{\rm B}=2r_{\rm C}$. We find that even a small difference between $z_{{\rm C},s}$ can lead to a large difference between $z_{{\rm B},s}$ when the void is deeper at the center. And a deeper void (or a smaller $\delta_{0,s}$) usually \Mov{favours} a wider void (or a larger $z_{{\rm B},s}$). Therefore, even a smaller cosmic inhomogeneity at the center of a deeper void can lead to a larger cosmic anisotropy at the boundary of this void.

\begin{table}
\captionsetup{justification=raggedright} \caption{The constraints on void parameters\MovE{ (see Sec.~\ref{sec:R})}
and the absolute magnitude.}
\label{tb:result} %
\begin{tabular}{cccc}
\hline 
${\rm Dataset}$  & $\delta_{0}~(68\%)$  & $z_{{\rm C}}~(95\%)$  & $M_{{\rm B}}~(68\%)$ \tabularnewline
\hline 
${\rm P_{{\rm All}}}$  & $-0.05\pm0.10$  & $<1.97$  & $-19.41\pm0.04$ \tabularnewline
\hline 
${\rm P_{{\rm N}}}$  & $-0.11\pm0.09$  & $<2.16$  & \multirow{2}{*}{$-19.40\pm0.03$ }\tabularnewline
${\rm P_{{\rm S}}}$  & $-0.07\pm0.09$  & $<2.05$  & \tabularnewline
\hline 
${\rm P_{{\rm W}}}$  & $-0.08\pm0.09$  & $<2.02$  & \multirow{2}{*}{$-19.40\pm0.03$ }\tabularnewline
${\rm P_{{\rm E}}}$  & $-0.11\pm0.09$  & $<2.28$  & \tabularnewline
\hline 
${\rm P_{{\rm NW}}}$  & $-0.11\pm0.08$  & $<2.16$  & \multirow{4}{*}{$-19.40\pm0.03$ }\tabularnewline
${\rm P_{{\rm NE}}}$  & $-0.09\pm0.09$  & $<2.10$  & \tabularnewline
${\rm P_{{\rm SW}}}$  & $-0.06\pm0.08$  & $<1.88$  & \tabularnewline
${\rm P_{{\rm SE}}}$  & $-0.15\pm0.08$  & $<2.34$  & \tabularnewline
\hline 
\end{tabular}
\end{table}

\section{Summary and Discussion}
\label{sec:SD}
In this paper, we try to test both of the local cosmic inhomogeneity and anisotropy on small scales at the same time using the SNIa data of Pantheon. Similar\Mov{ly} to the hemisphere comparison method, \Mov{however using} the LTB metric instead of the FRLW metric, we first divide the full dataset into several \Mov{data} subsets and then use the \Mov{data} subsets to constrain the void parameters $\{\delta_{0,s},z_{{\rm C},s}\}$ in the corresponding direction. Due to the paucity of data, we concentrate on the profile of the void at $z<z_{\rm C}$ where the curvature changes by $10\%$. \Mov{Despite this maneuver, only $\delta_{0,s}$ turns out well constrained, contrary to $z_{{\rm C},s}$}. The constraints on $\delta_{0,s}$ for all \Mov{division} cases are consistent with the FLRW metric at $95\%$ CL. The constraints on $z_{{\rm C},s}$ for all \Mov{division} cases are almost beyond $2.26$ at $95\%$ CL. That is to say, our constraints are too weak to challenge the cosmic homogeneity and isotropy. If the local cosmic void \Mov{does} exist, however, even a smaller cosmic inhomogeneity at the center of a deeper void can lead to a larger cosmic anisotropy at the boundary of this void.

Although our results are consistent with the cosmic homogeneity and isotropy at $95\%$ CL, \MovE{as the $\delta_{0,s}$ are consistent with 0 (denoting cosmic homogeneity) while the constraints on $\delta_{0,s}$ are consistent with each other (indicating cosmic isotropy), }there are some deviations from FLRW metric at $68\%$ CL. These deviations result \Mov{either }from the paucity of data in the subset or real physics in the corresponding direction. Therefore, more SNIa observations or other cosmological observations are needed to alleviate the \Mov{possible effect of }paucity of data in certain directions. 

\begin{acknowledgments}
We acknowledge the use of HPC Cluster of Tianhe II in National Supercomputing Center in Guangzhou. Ke Wang is supported by grants from NSFC (grant No. 12005084 and grant No.12247101) and grants from the China Manned Space Project with NO. CMS-CSST-2021-B01. MLeD acknowledges financial support by the Lanzhou University starting fund, the Fundamental Research Funds for the Central Universities (Grant No. lzujbky-2019-25), the National Science Foundation of China (grant No. 12047501), and the 111 Project under Grant No. B20063.
\end{acknowledgments}

\end{document}